\documentclass[aps,pre,twocolumn,groupedaddress,showpacs]{revtex4}
\usepackage{graphicx}
\usepackage{dcolumn}
\usepackage{bm}
\usepackage{amsmath}
\usepackage{epsfig}
\input epsf
\epsfclipon
\begin{document}
\title{Interplay between network structure and self-organized
criticality}
\author{Piotr Fronczak, Agata Fronczak and Janusz A. Ho\l yst}
\affiliation{Faculty of Physics and Center of Excellence for
Complex Systems Research, Warsaw University of Technology,
Koszykowa 75, PL-00-662 Warsaw, Poland}
\date{\today}

\begin{abstract}
We investigate, by numerical simulations, how the avalanche
dynamics of the Bak-Tang-Wiesenfeld (BTW) sandpile model can
induce emergence of scale-free (SF) networks and how this emerging structure affects dynamics of the system. We also discuss how the observed phenomenon can be used to explain evolution of scientific collaboration.
\end{abstract}


\maketitle


Since the discovery of self-organized criticality (SOC) by Bak,
Tang and Wisenfeld (BTW) \cite{bak}, the phenomenon has received
enormous attention among the researchers. During these almost
twenty years, dozens of variants of the original sandpile model
\cite{bak,sandpile} were studied
\cite{variant1,variant2,variant3,variant4} and a number of
examples of SOC in real world were discovered. One of the most
remarkable features that characterizes self-organized criticality
is the power law distribution of the characteristic events. The
feature combined with the abundance of real-world networks with
scale-free (SF) degree distribution \cite{0a,0b,0c,0d} may give rise to suspicion that
there exists mutual relation between the two issues. Although the
idea has been already mentioned in several papers (see for example
\cite{bianconi,paczuski}), according to our knowledge, so far no-one has
established a link between the two phenomena.

At the beginning let us recap the rules of sandpile model which is
a simple intuitive example of self-organized criticality. It is a cellular
automaton whose configuration is determined by the integer
variable $c_{i}$ (the height of the "sand column") at every node
$i$ of the network. Depending on network structure minor differences in definition can occur. Here we follow the definition of BTW model for random networks with a given degree distribution $p(k)$ \cite{goh1}. The dynamics is defined by the following
simple rules: A grain of sand is added at a randomly selected node $i$: $c_{i}
\rightarrow c_{i}+1$. A sand column with a height $c_{i}\geq k_{i}$, where $k_{i}$
is equal to degree of the node $i$, becomes unstable and collapses by
distributing one grain of sand to each of it's $k_{i}$ neighbors.
This may cause some of them to become unstable and collapse at the
next time step. This in turn can lead to an avalanche of next instabilities. During the
evolution a small fraction $f$ of grains is lost, which
prevents system to become overloaded. When
avalanche dies another grain of sand is added.

In the sandpile model distributions of avalanche sizes (measured as total number of
topplings in the avalanche), avalanche areas (the number of
distinct nodes participating in a given avalanche), avalanche
durations as well as many other statistics follow power law distributions.

Studies of sandpile dynamics carried out so far show that the characteristic exponents of measured distributions depend on the network topology. For example the avalanche size exponent is $\tau=1$ for $2D$ square lattice \cite{bak} and $\tau=1.33$ for $3D$ cubic lattice \cite{bak}. In Erd\H{o}s-R\'{e}nyi (ER) random networks $\tau=1.5$ \cite{bonabeau}. Recently, Goh et al. have studied sandpile dynamics on scale free networks $p(k)\sim k^{-\gamma}$ \cite{goh1,goh2} and they have shown that the avalanche area exponent $\tau$ is independent on the average network connectivity $\langle k\rangle$ and changes with the exponent $\gamma$ of the degree distribution. They have obtained 
\begin{equation}
\tau=\frac{\gamma}{\gamma-1}
\label{eq_goh}
\end{equation}
 in the range $2<\gamma<3$ and $\tau=1.5$ for $\gamma >3$. The question we ask in the paper is the following: how the avalanche area exponent behaves when the network topology depends on sandpile dynamics, i.e. when there are mutual interactions between the network structure and network dynamics?

In \cite{bianconi} Bianconi and Marsili proposed a simple model in
which the network reorganizes its structure as a consequence of
avalanches of rewiring processes. The only parameter of the model
which influences the rewiring and in consequence the network
structure is a type of probability that a chosen node becomes
unstable and has to be rewired. Choosing this probability as a power law one can set the system in a critical
state and force the network to take a power law degree distribution. 

In the present paper, instead of forcing the network to stay in a critical state, we allow the system to naturally evolve towards the critical region. In our model:
\begin{enumerate}
\item the degree distribution of the considered networks changes due to the distribution of sandpile avalanches on this network and
\item the avalanche size distribution changes because the network structure evolves.
\end{enumerate}
These two mechanisms influence each other and lead to the equilibrium
point in which the shapes of avalanche distribution and degree
distribution become identical. In the second part of the paper we
propose how the presented phenomenon can be applied to modelling of
evolution of scientific collaboration.

In order to complete the rules of our model, apart from the rules of sandpile model recapitulated above, we define 
the rewiring process in the following way: each end of a link has been
assigned a value specifying the time when it was rewired for the last time. After an avalanche of area $A$, the number of $A$ 'oldest' ends of links are rewired to the node which has triggered the avalanche off.

\begin{figure}
{\epsfig{file=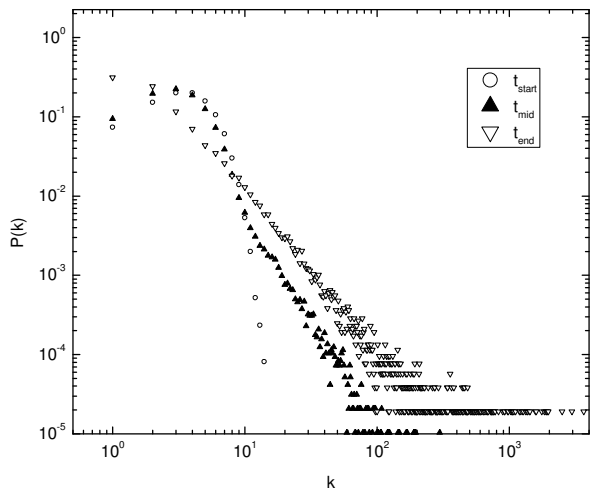,width=.9\columnwidth}} \caption{Node degree
distributions of rewired network in three time moments:
$t_{start}=0$, $t_{mid}=2500$ and $t_{end}=10000$.} \label{fig_2}
\end{figure}

\begin{figure}
{\epsfig{file=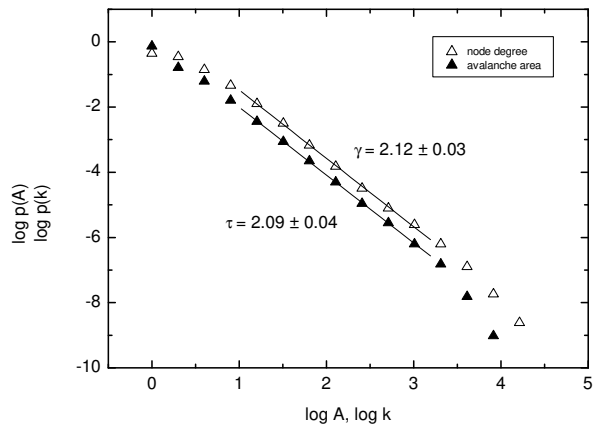,width=.9\columnwidth}}
\caption{Distributions of avalanche area and node degree in time
$t_{end}$. Data are logarithmically binned. Lines are linearly
fitted with the values indicated at the figure.} \label{fig_3}
\end{figure}

In our studies all networks have:  $\langle k\rangle=4$, number of nodes $N=10^{5}$ and
$f=10^{-4}$. We start our simulation with Erd\H{o}s-R\'{e}nyi random network (which corresponds to
$\gamma=\infty$ in static SF networks). Time unit that has been used was simply one
avalanche and we have carried out our simulations for $t=10000$ steps. Fig. \ref{fig_2} presents three
snapshots of the node degree distribution in three different moments:
$t_{start}=0$, $t_{mid}=2500$ and $t_{end}=10000$. Comparing the snapshots, one can see that the
network reorganizes itself from Poissonian, through a mixture
of Poissonian and scale-free and finally settles on a pure
scale-free degree distribution with the well established
characteristic exponent $\gamma\cong 2$. At the same time the characteristic exponent of the avalanche area distribution increases from $\tau=1.5$, which is the known result for ER random networks \cite{bonabeau}, to $\tau=2$. As one can see at Fig. \ref{fig_3} in the course of simulation the two exponents converge to the common equilibrium value $\tau=\gamma=2$. 

\begin{figure}
{\epsfig{file=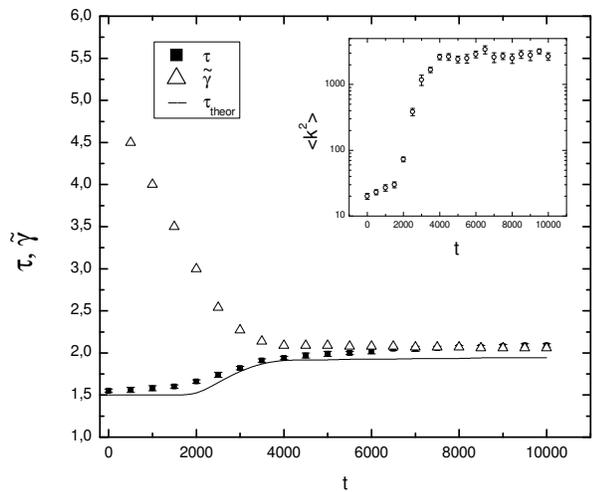,width=.9\columnwidth}}
\caption{Process of equilibration of exponents $\tilde{\gamma}$ and $\tau$. Solid line presents theoretical $\tau_{theor}$ obtained from $\tilde{\gamma}$ and eq. \ref{eq_goh}. Inset: second moment of the degree distribution in time.}
\label{fig_4}
\end{figure}

\begin{figure}
\epsfig{file=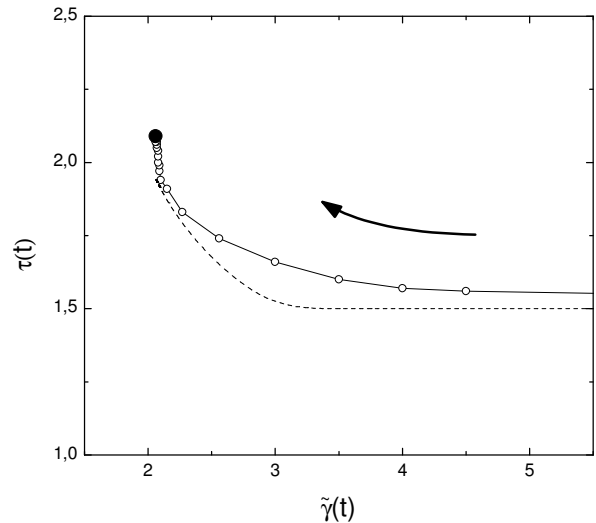,width=.9\columnwidth} \caption{Dependence of
avalanche area exponent $\tau$ on parameter $\tilde{\gamma}$ of
generated scale-free network. The arrow shows direction of a movement in space of parameters during the process of equilibration. Measurements are done in equal time steps $\Delta t=500$ and marked as open circles. The black dot depicts the fixed point of the process. Dashed line presents theoretical $\tau_{theor}$ obtained from $\tilde{\gamma}$ and eq. \ref{eq_goh}.} \label{fig_1}
\end{figure}

Fig. \ref{fig_4} presents the convergence process in a more detailed way. We define there a new parameter $\tilde{\gamma}(t)$ that in some sense may be understood as the characteristic exponent of fat-tailed degree distributions and may be compared to $\gamma$. $\tilde{\gamma}(t)$ is simply obtained from the second moment of the degree distribution that is know from simulations. Given $\langle k^{2}\rangle$, we numerically solve the below equation for $\tilde{\gamma}(t)$ 
\begin{equation}
\langle k^{2}\rangle=\sum_{k=1}^{N}k^{2}p_{an}(k)
\label{k2}
\end{equation}
where 
\begin{equation}
p_{an}(k)=\langle k \rangle \frac{(\tilde{\gamma}-2)^{\tilde{\gamma}-1}}{(\tilde{\gamma}-1)^{\tilde{\gamma}-2}}
\frac{\Gamma(k-\tilde{\gamma}+1,\langle k \rangle \frac{\tilde{\gamma}-2}{\tilde{\gamma}-1})}{\Gamma(k+1)}
\label{static}
\end{equation}
is the known analytic solution of the static model \cite{analstaticmodel}.
The new parameter has been introduced because in intermediate times of the simulation the degree distribution does not follow a pure power law. At the mentioned figure one can see that the value of the exponent $\tilde{\gamma}$ (open triangles) decreases from $\infty$ to $2.1$.  Simultaneously, the parameter $\tau$ characterizing the sandpile dynamics (solid squares) increases from $1.5$ and finally settles at the equilibrium value of $2.1$. The values of $\tau$ fairly good agree with the values of $\tau_{theor}$ (solid line) calculated from the relation (\ref{eq_goh}) derived for static SF networks by Goh et al. \cite{goh1}. The last observation let us suspect that during simulation the system moves along the trajectory given by the formula (see Fig. \ref{fig_1})
\begin{equation}
\tau(t)=\frac{\tilde{\gamma}(t)}{\tilde{\gamma}(t)-1},
\label{eq_pf}
\end{equation}
and respectively the equilibrium point may be calculated from the above equation when one assumes  $\tilde{\gamma}=\tau$. 

The exponents $\tilde{\gamma}=\tau=2$ characterizing the final critical state seem to be robust against different threshold assignment strategies in sandpile dynamics. In order to support the last statement let us mention two papers \cite{goh2,paczuski} in which we have found probable symptoms of such a universality. In \cite{goh2}, a class of sandpile models is studied. In this class the threshold height of a node is set as $k^{1-\eta}$, where $0\leq\eta<1$ is a parameter of the class. The avalanche size exponent is received as $\tau=(\gamma-2\eta)/(\gamma-1-\eta)$. If one assumes that due to self-organization and rewiring process in our model $\tau=\gamma$, then one obtains $\tau=\gamma=2$ independently on $\eta$. The second example is a model of rapid rearrangements in the network of the magnetic field flows in the Sun corona \cite{paczuski}. The authors show that the avalanches of link reconnections and scale free structure of the considered network co-generate each other. They also show that for the equilibrium the degree distribution exponent $\gamma=2$. Unfortunately, they do not present reconnection distribution exponent which corresponds to $\tau$. 

In fact, the precise value of the fixed point is a bit larger than
$2$, about $2.1$. It can result from the finite size effects
or from the fact, that in the vicinity of $\gamma=2$ the considered networks
are highly correlated but the way we perform rewiring
includes a small random contribution (it is known that in this
range of parameters the network should be correlated
disassortatively, so instead of rewiring the oldest end of the
link we should perhaps rewire the least disassorative link).

In the last part of the paper we would like to show how the
observed phenomenon can be used to construct a simple model of
evolution of scientific connections. 

In the model each node represents a scientist. A link between two
scientists represents the fact that they can exchange new ideas
and draw inspiration from each other. A scientist has
its own potential (hidden variable) which describes his/her ability to
produce an interesting paper. The potential is a non decreasing
function of time (we rather collect ideas, do not lose them). If
the potential crosses a critical level then an interesting
publication is born and the potential is reduced to zero i.e. the
scientist has to collect new ideas from the beginning. From time
to time an avalanche of such interesting publications occurs. All
those publications use some ideas and draw inspiration from the
first one in the avalanche and of course cite it. After the avalanche
the scientist who triggered the avalanche off attracts attention of
other scientists. They will try to establish a closer
collaboration with such a famous person. They will be glad to
present to him/her their best ideas and they will follow his/her next
papers because they expect that it is worth to read them. The larger
avalanche the scientist caused the larger number of people will be
interested in his/her publications. Because it is
impossible to observe works of all people every scientist reads
only the papers of a group of the most important persons. That is
why after finding a new person who become interesting, the
scientist has to resign from observation of other less interesting
one.

In the model that has been presented above one can find mechanisms which can be
modeled by the phenomenon described in the first part of the paper. The potential may be considered as
a column of grains and every grain represents a quantum of idea. To make the potential equal for all scientists one can normalize it by a scientist's degree.
By an avalanche we understand an occurrence of some scientific
sub-domain (like physics of complex networks or self-organized criticality) so
the timescale of such an avalanche will be expressed in years.

The above model just gives an idea where the
presented phenomenon can be found. However, since the term 'collaboration'
does not only mean common papers but every manifestation of
exchange of ideas (appeared as co-authorship, citing and even
common discussions), the applicability of the model to real data may suffer a number of difficulties.

To conclude, in this paper we have presented by numerical
simulations how the avalanche dynamics of the Bak-Tang-Wiesenfeld
sandpile model and the network structure may influence each
other. Such an interplay between dynamics and structure leads to
self-organization in which the shapes of avalanche distribution
and degree distribution become identical. We suspect that the value of
both exponents $\gamma=\tau=2$ may be universal for a large class of SOC phenomena
in which the critical behavior occurs not 'on' the network structure but 'in' the structure. We also show how the observed phenomenon
can be used to construct a simple model of evolution of
scientific collaborations.

\begin{acknowledgments}
We are thankful to V.M. Eguiluz for interesting discussions. The
work has been supported by European Commission Project CREEN
FP6-2003-NEST-Path-012864. A.F. acknowledges financial support from the Foundation for Polish Science (FNP 2005). A.F. and J.A.H. were partially supported by the State Committee for Scientific Research in Poland (Grant No. 1P03B04727).
\end{acknowledgments}


\begin{thebibliography}{99}
\bibitem{bak} P. Bak, C. Tang, and K. Wiesenfeld, Phys. Rev. Lett. {\bf 59}, 381 (1987).
\bibitem{sandpile} P. Bak, C. Tang, and K. Wiesenfeld,Phys. Rev. A {\bf 38}, 364 (1988).
\bibitem{variant1} D. Dhar, Phys. Rev. Lett {\bf 64}, 1613 (1990).
\bibitem{variant2} D. Dhar and R. Ramaswamy, Phys. Rev. Lett. {\bf 63}, 1659 (1989).
\bibitem{variant3} H.J. Jensen, {\it Self-Organized Criticality: Emergent Complex Behaviour in Physical and Biological Systems}, Cambridge University Press, Cambridge, 1998.
\bibitem{variant4} C. Maes, F. Redig and E. Saada, Ann. Probab. {\bf 30}, 2081 (2002).
\bibitem{0a} S.Bornholdt and H.G.Schuster, {\it Handbook of Graphs and
networks}, Wiley-Vch (2002).
\bibitem{0b} S.N. Dorogovtsev and J.F.F.Mendes, {\it Evolution of
Networks}, Oxford Univ.Press (2003).
\bibitem{0c} R.Albert and A.L.Barab\'asi, Rev. Mod. Phys. {\bf 74} 47 (2002).
\bibitem{0d} S.N.Dorogovtshev and J.F.F.Mendes, Adv.Phys. {\bf 51} 1079 (2002).
\bibitem{bianconi} G. Bianconi and M. Marsili, Phys. Rev. E {\bf 70}, 035105(R) (2004).
\bibitem{paczuski} M. Paczuski and D. Hughes, Physica A {\bf 342}, 158 (2004).
\bibitem{bonabeau} E. Bonabeau, J. Phys. Soc. Japan {\bf 64}, 327 (1995).
\bibitem{goh1} K.I. Goh, D.S. Lee, B. Kahng and D. Kim, Phys. Rev. Lett. {\bf 91}, 148701 (2003).
\bibitem{goh2} D.S. Lee, K.I. Goh, B. Kahng and D. Kim, Physica A {\bf 338}, 84 (2004).
\bibitem{static} K. I. Goh, B. Kahng and D. Kim, Phys. Rev. Lett. {\bf 87}, 278701 (2001).
\bibitem{analstaticmodel} M. Catanzaro and R. Pastor-Satorras, Eur. Phys. J. B {\bf 44}, 241 (2005).


\end{thebibliography}
\end{document}